\documentclass[runningheads]{llncs}
\usepackage{graphicx}
\usepackage{todonotes}

\usepackage{CJKutf8}
\begin{document}
\title{How We Express Ourselves Freely: Censorship, Self-censorship, and Anti-censorship on a Chinese Social Media}
\titlerunning{Censorship, Self-censorship, and Anti-censorship on a Chinese Social Media}
%
\author{Xiang Chen\inst{1}\orcidID{0000-0002-0573-9651} \and
Jiamu Xie\inst{2}\orcidID{0000-0001-6345-1886} \and\\
Zixin Wang\inst{3}\orcidID{0000-0002-8594-3803}\and
Bohui Shen\inst{4}\orcidID{0000-0003-1409-8822}\and\\
Zhixuan Zhou\thanks{Corresponding author.} \inst{5}\orcidID{0000-0001-9388-4156}}
\authorrunning{X. Chen, J. Xie et al.}
%
%
\institute{King Abdullah University of Science and Technology, Saudi Arabia
\email{xiang.chen@kaust.edu.sa}\\ 
\and
Wuhan University, China\\
\email{2019300001053@whu.edu.cn}\\
 \and
Zhejiang University, China\\
\email{zixin\_wang@zju.edu.cn}
 \and
BNU-HKBU United International College, China\\
\email{zosh.olivia@gmail.com}
 \and
University of Illinois at Urbana-Champaign, United States\\
\email{zz78@illinois.edu}
}
\maketitle              
\begin{abstract}
Censorship, anti-censorship, and self-censorship in an authoritarian regime have been extensively studies, yet the relationship between these intertwined factors is not well understood. In this paper, we report results of a large-scale survey study ($N=526$) with Sina Weibo users toward bridging this research gap. Through descriptive statistics, correlation analysis, and regression analysis, we uncover how users are being censored, how and why they conduct self-censorship on different topics and in different scenarios (i.e., post, repost, and comment), and their various anti-censorship strategies. We further identify the metrics of censorship and self-censorship, find the influence factors, and construct a mediation model to measure their relationship. Based on these findings, we discuss implications for democratic social media design and future censorship research.

\end{abstract}
\section{Introduction}
Sina Weibo is a Chinese social media platform providing micro-blogging services, which can be thought of as a Twitter equivalent \cite{weibo}. It provides a medium for people to participate in civic life \cite{politics} and express themselves \cite{anonymous}. Weibo has a huge user base: according to a financial report published by Sina in 2021, the average daily active users of Weibo in December 2021 reached 249 million \cite{weibo:data}. The implication of the large user base is that the dark side of social media, e.g., fake news \cite{fake:news} and discrimination speech \cite{ethics}, can potentially affect a large population.  

Nationalism and censorship of political discussion on Weibo has long been reported \cite{nationalism}, which negatively affects people's freedom of speech. Research has been conducted to understand users' attitudes toward censorship \cite{censorship:2}, anti-censorship strategies emerging from the user community \cite{censorship:5}, and self-censorship of media \cite{media:censorship}. However, there is a lack of understanding on how these intertwined factors interact with each other. 

In this paper, we report the results of a survey study ($N=526$) conducted with Sina Weibo users, toward understanding censorship, self-censorship, and anti-censorship practices in the current Chinese social media landscape from a user perspective, as well as their relationships.

Through the descriptive statistics, we construct the profile of Weibo users based on their demographic information, including age range, occupation, and the time and frequency of Weibo usage. We also discover the ways and topics that Weibo users are usually censored on, their self-censorship practices, and anti-censorship strategies commonly used by the respondents. Based on principal component analysis (PCA), we construct three metrics to measure the attitudes of Weibo users towards censorship and anti-censorship, namely, perceived necessity of self-censorship, impact of self-censorship on users’ expression desire and mood, and support for censorship. Through correlation analysis, we find the influence factors significantly correlated to these three metrics. On top of this, we build an intermediate model for these three metrics and show the mediation effect of censorship and self-censorship: the perceived necessity of self-censorship will increase the impact of self-censorship on users’ expression desire and mood, which in turn will increase support for censorship.

Thus our contributions are two-fold. First, we update previous understanding of the status quo of the censorship infrastructure and situation on Chinese social media. Second, we approach the relationship between censorship, self-censorship, and anti-censorship from a user perspective. 

In the sections below, we will first discuss literature on censorship, anti-censorship, and self-censorship. Then we describe our method and results, followed by a discussion of main takeaways from this study and design implications for more democratic social media in an authoritarian regime.

\section{Related Work}

\subsection{Censorship}
Internet censorship has been extensively studied, especially in the context of authoritarian regimes \cite{censorship:xu}. Twitter has a standard interface for government agencies to request that individual tweets or accounts be censored. Tanash et al. discovered over a quarter million censored
tweets out of 20 million from late 2014 to early 2015 in Turkey \cite{turkey}. Most of the censored tweets were found to contain political content, often criticisms of the Turkish government. Twitter under-reported the number of censored tweets in Turkey, which might hold for other countries as well. The situation of social media censorship in China can be more unknown. The platforms are known to be closely regulated by the Chinese government and asked to censor political content, yet they never disclose how many posts or accounts they censor each year --- it is a black-box. However, by applying a statistical analysis on massive social media data of Sina Weibo and Twitter, Bamman et al. uncovered a set a politically sensitive terms whose presence in a message could lead to anomalously higher rates of deletion \cite{deletion}.

It is argued that censorship, when detected by citizens, will have an adverse impact on their assessment of the government, because censorship signals the government's inability to address the issue being censored \cite{censorship:1}. However, perceived government intrusion is strongly correlated with privacy concern and self-protective behavior, and trust towards other Weibo participants has a significant negative relationship with self-protective behavior \cite{censorship:3}. Generally, Internet users have varied attitudes and perceptions of censorship. Users' demographic backgrounds, Internet usage experience, and personality are known to influence their attitudes toward censorship. Those with an authoritarian personality tend to support censorship \cite{censorship:2}. One relevant paper uncovered common topics when Chinese users discussed censorship on Twitter, which has long been blocked by the Chinese government, namely, sharing technical
knowledge, expressing political opinions, and disseminating alternative news items \cite{Chinese:Twitter}. While these Chinese users either reside overseas or are able to use VPN to access Twitter, we sought out to understand how people blocked by the Great Firewall \cite{great:firewall} perceived censorship. 

\subsection{Anti-censorship}
It is possible to predict if a social media post will be censored. For example, Ng et al. used linguistic properties of social media posts to automatically predict if they were going to be censored \cite{censorship:4}. Algorithmically bypassing the censorship of social media is also possible. Hiruncharoenvate et al. presented a non-deterministic algorithm for generating homophones that created large numbers of false positives for censors, making it difficult to locate banned conversations \cite{homophone}. 

From a user perspective, though they have limited knowledge of natural language processing, Weibo users are found to intuitively express in machine unreadable ways, e.g., image-based content, to resist the censorship infrastructure \cite{censorship:5}. Chinese users have also adopted variants of words, i.e., morphs, to avoid keyword-based censorship \cite{morph}. Here we use a larger-scale survey study to uncover more strategies used to resist the censorship infrastructure imposed by Weibo.

\subsection{Self-censorship}
It is common for social media users to exhibit some level of last-minute self-censorship of their posts, mostly for the sake of the ``perceived audience'', and people with more boundaries tend to self-censor more \cite{self:censorship}. At a larger scale, media self-censorship in China has been found to increase the possibility of the publication of reports on highly politically sensitive topics \cite{media:censorship}. In this paper, we focus on the latter purpose of self-censorship, i.e., avoid being censored. Specifically, we look into how users experience self-censorship, and how self-censorship practices are associated with censorship. 

The most self-censored regimes in China include discussions of LGBTQ \cite{lgbtq} and political issues \cite{political:discussion}, and release of music \cite{music}. 
In such cases, self-censorship is not only about how to circumvent censorship, but also about people's efforts to negotiate with the authorities \cite{music}. 

While censorship can potentially inspire self-censorship or anti-censorship practices, the relationship between these factors have hardly been discussed in previous literature. We bridge this research gap with this current survey study.

\section{Method}

\subsection{Survey Flow}
Through this study, we aimed to understand user attitudes toward censorship, self-censorship, and anti-censorship, as well as relationships between these factors. To this end, we designed a survey comprised of 38 questions. Below, we elaborate on each section of the survey.

\subsubsection{Introduction and Screening.} We started by introducing the concept of self-censorship. In our definition, self-censorship is ``Internet users' inspection and examination of their own speech. Choices after self-censorship can be whether or not to say something, what to say, and how to say.'' 

We filtered participants with two screening questions: (1) whether one has used Weibo, and (2) whether one has practiced self-censorship.

\subsubsection{Demographics and Weibo Usage.} To know more demographic information of the survey respondents, we asked questions regarding gender, age, occupation, location (province), yearly income, party membership, educational background, and relevant study/work experience in computer/software. We gave participants the option not to disclose party membership, given the potentially sensitive nature of this question. Though understanding the effect of demographic information on people's self-censorship choices was only a secondary goal of our investigation, it could nevertheless situate our analysis in a specific cultural context. 

Since we aimed to investigate Weibo users' self-censorship choices, we also needed to ask about their usage of Weibo, including the amount of time spent on Weibo via computers (desktop and laptop) and mobile devices (mobile phone, iPad, etc.), years of experience with Weibo, usage frequency, activities on Weibo (posting, commenting, news feed, etc.), and number of followers and following.

\subsubsection{Self-censorship.} We explicitly asked participants' experience, purpose, and perspective of self-censorship, and used a five-point Likert scale (``Strongly Disagree'' to ``Strongly Agree'') to probe their attitude toward self-censorship.

We further used three five-point Likert scales to identify participants' choices of self-censorship when posting, commenting and reposting, respectively, on Weibo. We differentiated between self-censorship choices of different topics, namely, entertainment, sports, current events, daily life, covid, humor, video games, and finance. Some of the listed topics are potentially more susceptible to self-censorship as well as censorship, such as current events and covid.  

\subsubsection{Censorship and Anti-censorship.} To identify the effect of users' experiences/attitudes of censorship on self-censorship choices, we asked participants' experiences, topics and ways of being censored by Weibo, as well as common anti-censorship strategies used by Chinese people. The research team, which consisted of five researchers from China, identified a wide range of anti-censorship strategies based on their years of experience of Weibo usage. We also used a five-point Likert scale to probe participants' attitude toward censorship.

\subsubsection{IP Address.} Recently, Weibo, as well as some other social media in China, started to reveal users' IP address on their homepage and comments, possibly to crack down on discussion surrounding pandemic policy and other sensitive political topics. To identify the effect of the disclosure of IP address on users' self-censorship choices, we asked participants whether they would more strictly self-censor their speech after the change. We also asked perceived pros and cons of, and attitudes toward the revealing of IP address. Further, a user's willingness to pay for a feature can be used as a proxy for how much the user values that feature. With this in mind, we asked how much the participants were willing to pay in order to hide their IP address. 

\subsection{Data Collection}
We recruited participants on different social media platforms in China, including WeChat and Weibo itself. Personal contacts of the authors were also requested to distribute the survey link. 

Initially, we used Wenjuanxing\footnote{\url{https://www.wjx.cn/}}, a Chinese survey design and distribution platform, to facilitate the data collection process. However, after two days, our survey was stopped and recycled by the platform, stating that our survey contained politically sensitive questions and information. Notably, all survey platforms based in China were overseen by the Chinese government, as stated in their terms of use, and one needed to use her real identity (i.e., ID card) to distribute surveys via these platforms. Thus, we alternatively used Qualtrics\footnote{\url{https://www.qualtrics.com/}} to make the survey distribution possible.

When one researcher, who had about 3,000 followers on Weibo at the time of this study, distributed the survey, heated discussion took place in the comments. Participants were enthusiastic about our topics of investigation, and over 300 surveys were filled in the first few hours. They complained that they had long suffered from censorship of Weibo, and were even unaware of their own anti-censorship or self-censorship practices. Due to the wide spread of this survey, unfortunately, the Weibo account of this researcher was suspended. After a week, the account of another researcher who had around 2,000 followers was also suspended for distributing the survey.

We gained confidence in the ethical aspect of our research and our personal safety by finding a large body of literature on quantitatively or qualitatively researching censorship practices in China.

\section{Results}
In the end, a total of 1,346 responses were received, but many of them were missing data or unfinished. After filtering them out, we had 526 (39\%) valid questionnaires. Among them, 523 (99.4\%) respondents answered they have used Weibo, while 3 (0.6\%) answered they have not used Weibo, and the subsequent analysis only considered those who had used Weibo.

\subsection{Descriptive Statistics}
\subsubsection{Demographic Information}
Among the survey respondents, 18\% of them were male, and 80\% were female. Although the non-binary option was given in our questionnaire, only 2\% selected it. The large majority of survey respondents were young people: people aged 18-24 accounted for 70\%, and people aged 25-34 accounted for 25\% of the whole population. Students accounted for 68\% of the survey respondents. 
Since our participants were mostly students, 59\% of them reported to have no income. 10\% of the survey respondents had party membership, and 51\% of them were members of the Communist Youth League. Some noted that they ``automatically became Communist Youth League members in junior high school'', and were ``annoyed with the identity''. Interestingly, when asked where they lived, some participants chose not to disclose, because ``the revealing of IP address on Weibo has made them lose a sense of safety.'' People with and without computer related study/work experience were evenly distributed.

\subsubsection{Weibo Usage}
Participants preferred using mobile devices (3.0 hours on average per day) to use Weibo rather than computers (1.7 hours on average per day). They have used Weibo for 3.57 years on average, and 66\% of them had more than five years of experience using Weibo, identifying as experienced users. The majority of our participants were also frequent Weibo users, with 89\% of them using Weibo multiple times a day. Weibo was mainly used to post (80\%), comment (68\%), watch news (74\%), make friends (30\%), and following idols (31\%).

Many of our participants had a relatively large fan base. 47\% of them had more than 100 followers. Thus they might see a need to be mindful of what they post.

\subsubsection{Self-censorship}
Among those who ever used Weibo, 90.4\% acknowledged that they had practiced self-censorship. Averaged over 3 Likert-scale questions regarding user attitudes toward self-censorship, it turned out that the participants diverged in their perception of the necessity of self-censorship. However, they largely agreed that self-censorship decreased their willingness to express (84\%), and that self-censorship negatively affected their mood (80\%).

Participants practiced self-censorship to bypass the censorship (57\%), avoid being deleted (53\%), avoid being suspended (61\%), either temporarily or permanently, avoid being reported or cyberbullied (43\%), and avoid being summoned by the police (48\%). Other self-reported motivations for self-censorship included fear of conflicts, avoiding being noticed by acquaintances in real life, self-presentation, avoiding privacy leakage, habit, etc.

\begin{figure}[htp]
	\centering
	\includegraphics[width=0.99\linewidth]{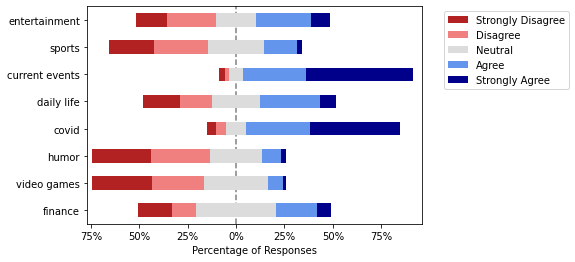}
	\vspace{-0.5cm}
	\caption{Tendency to self-censor on different topics when users \textbf{post}.}
	\label{fig:1}
	\vspace{-0.5cm}
\end{figure}

\begin{figure}[htp]
	\centering
	\includegraphics[width=0.99\linewidth]{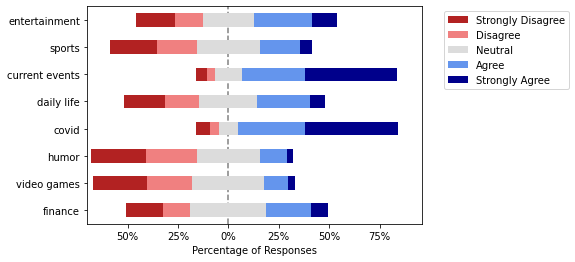}
	\vspace{-0.5cm}
	\caption{Tendency to self-censor on different topics when users \textbf{comment}.}
	\label{fig:2}
	\vspace{-0.5cm}
\end{figure}

\begin{figure}[ht]
	\centering
	\includegraphics[width=0.99\linewidth]{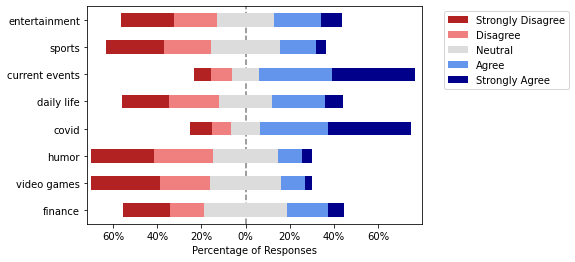}
	\vspace{-0.5cm}
	\caption{Tendency to self-censor on different topics when users \textbf{repost}.}
	\label{fig:3}
	\vspace{-0.5cm}
\end{figure}

Figure \ref{fig:1}, Figure \ref{fig:2}, and Figure \ref{fig:3} show the user's self-censorship practices in different usage scenarios and different topics. Users tend to practice more self-censorship when it comes to current events and pandemic, and less self-censorship on such topics as humor and video games. Figure \ref{fig:2} and Figure \ref{fig:3} reveal that users' self-censorship on reposting is more common than on commenting. One possible reason is that reposting will be displayed on their homepage, which is essentially similar to posting, while commenting is made under other users' Weibo posts, which will not cause much damage to their own Weibo accounts.



\subsubsection{Censorship}

79\% of Weibo users indicated that they had been censored, and the censored topics are mostly current events (79.0\%) and covid (69.0\%).
 
Censorship can imply different processes and consequences for users and we list some commonly mentioned ones below.
\begin{enumerate}
    \item The Weibo post cannot be sent;
    \item The repost button is removed after posting, and the content of this Weibo post can only be seen by one's followers or even oneself;
    \item The post is deleted after a while, due to either  moderation or other users' reporting; 
    \item The account is suspended, either being muted for a few days/weeks/months, or being permanently deleted;
    \item The user is summoned by the police or even put into prison.
\end{enumerate}

\subsubsection{IP Address}

In March, Weibo has begun to reveal users' IP locations when they post and comment, and on their account homepages, to ``discourage bad behavior''~\cite{IPAddress}. This feature cannot be proactively turned off by the users.

After the IP address was displayed on their Weibo homepage and comments, nearly half of the respondents indicated that their self-censorship efforts had not changed significantly, and they would pay as much attention to self-censorship as they do now; while less than half of the respondents indicated that they would increase the extent of self-censorship.

The vast majority (81.8\%) have a negative opinion regarding the act of revealing the IP address, but most (57.2\%) respondents said they would continue to use Weibo despite this. Another 31.7\% said that they did not know whether they would continue to use Weibo in the future.

\subsubsection{Anti-censorship Strategies}
Due to the increasing censorship, Weibo users have adopted or created the following strategies to avoid censorship of their published posts.
\begin{enumerate}
    \item Pinyin: phonetic transcription of mandarin Chinese;
    \item Acronym: initials of Pinyin;
    \item Homophone: substituting Chinese characters with the same or similar pronunciation;
    \item English translation;
    \item Martian language: characters that look similar, composed of non-normalized character symbols such as uncommon characters, split parts of Chinese characters, or many other unicode symbols;
    \item Emoji;
    \item Code name, e.g., addressing pandemic personel as ``Big White'' who are often in white coveralls;
    \item Mixing symbols among Chinese characters;
    \item Reversing or shuffling the order of Chinese characters;
    \item Satire; 
    \item Converting text to images;
    \item Adding graffiti or occlusion on converted images;
    \item Rotating or flipping the converted images;
    \item Posting an irrelevant yet safe Weibo, and put what they really want to say in the comment or in the edit log.
\end{enumerate}

The three anti-censorship strategies perceived as most effective by the respondents are satire, acronym, and homophones, which are most likely to bypass NLP-based or human moderation.


\subsection{A Medication Model Connecting Censorship and Self-censorship}
\subsubsection{Identifying the Metric of Censorship and Self-Censorship}\label{4.3.1}
Two groups of questions in our survey asked respondents about their attitudes toward censorship and self-censorship, respectively, and were quantified using the Likert scale. 

The three questions regarding users' perception of self-censorship are listed below.
\begin{enumerate}
\item Self-censorship is necessary for social media use;
\item Self-censorship reduces my willingness to express myself; 
\item Self-censorship negatively affects my mood.
\end{enumerate}
We use the Cronbach's $\alpha$ coefficient~\cite{cronbach1951coefficient} to measure the strength of consistency of these 3 questions.
Practically, a Cronbach's $\alpha$ of 0.7 or higher is considered acceptable. The standardized Cronbach's $\alpha$ coefficient of these 3 questions is 0.699, and when Question 1 is not taken into account, the coefficient of the remaining two questions rises to 0.749. Therefore, we suppose Questions 2 and 3 jointly reflect the impact of self-censorship on users' expression desire and mood, while Question~1 alone reflects the perceived necessity of self-censorship. Thus the two metrics of self-censorship are (perceived) \textbf{necessity of self-censorship} and \textbf{impact of self-censorship} on expression desire and mood. There will be scenarios like this: although self-censorship greatly affects users' mood, in order to be able to post, the users still think it necessary to practice self-censorship.

Similarly, several questions are used to probe users' attitudes toward censorship:
\begin{enumerate}
\item Weibo censorship is good for improving the Internet environment;
\item I would report Weibo posts holding a different opinion from mine;
\item I hope Weibo would remove the censorship mechanism;
\item Censorship shows no respect for freedom of speech;
\item I support suspending accounts disseminating extreme speech;
\item Censorship negatively affects my user experience with Weibo.
\end{enumerate}
Likewise, we use the Cronbach's $\alpha$ coefficient to measure the strength of consistency of these 6 questions. The coefficient is 0.824, which shows a strong consistency. We further calculate the Corrected Item-Total Correlation (CITC)~\cite{zijlmans2019item}. The CITC of Question 2 and 4 is 0.348 and 0.497, respectively, which indicates that the relationship between these two questions and the rest of the items is weak \cite{clark2019constructing}. Thus we did not take these two questions into account. This is also confirmed in the principal component analysis (PCA). The result of PCA shows that if all the 6 questions are combined into one principal component, the contributions of Questions 2 and 4 are very small. After removing them, the four questions 1, 3, 5, and 6 can be combined into one principal component, which can be regarded as the metric of \textbf{support for censorship} for further exploration.

\subsubsection{Correlation Test}
After identifying the metrics to measure self-censorship and censorship, we conduct the correlation analysis and hypothesis testing.

We performed t-test and analysed the variance of the correlations between the three metrics ($N$: the necessity of self-censorship, $I$: the impact of self-censorship on users' expression desire and mood, $S$: support for censorship) elaborated on above, as well as other variables and influencing factors in the questionnaire.

At the significance level of $p=0.01$, multiple variables show a significant correlation with these metrics, as seen in Table~\ref{tab-corr}. Motivations of self-censorship, e.g., avoiding account suspension, and avoiding being summoned by the police, are correlated to all three metrics. Censorship/self-censorship/anti-censorship experiences, i.e., having conducted self-censorship on Weibo, having been censored on Weibo, and having used anti-censorship strategies, are also correlated to all three metrics. 

The correlation of these three metrics themselves is shown in Table~\ref{tab-index}. As we can see, all three metrics are significantly correlated at the p = 0.01 level.

\begin{table}[ht]
\centering
\caption{Influence factors significantly correlated with the metrics of censorship and self-censorship ($p=0.01$).}\label{tab-corr}
\resizebox{\textwidth}{!}{%
\begin{tabular}{cc}
\hline
Metrics of Censorship/Self-Censorship &  Influence Factors \\ \hline
$I$: Impact of Self-censorship &  \begin{tabular}[c]{@{}l@{}}Avoid account suspension\\Avoid being summoned by the police\\Gender\\Age\\How long they have used Weibo\\Have conducted self-censorship on Weibo\\Have been censored on Weibo\\Have used anti-censorship strategies\end{tabular}\\ \hline
$N$: Necessity of Self-censorship &  \begin{tabular}[c]{@{}l@{}}Make sure Weibo posts can bypass the censorship\\Keep Weibo posts from being deleted\\Avoid account suspension\\ Avoid being summoned by the police\\Use Weibo to post\\Have conducted self-censorship on Weibo\\Have been censored on Weibo\\Have used anti-censorship strategies \end{tabular}\\ \hline
$S$: Support of Censorship & \begin{tabular}[c]{@{}l@{}}Make sure Weibo posts can bypass the censorship\\Keep Weibo posts from being deleted\\Avoid account suspension\\Avoid being summoned by the police\\Computer or software related experience\\Frequency of Weibo usage\\Use Weibo to post\\Have conducted self-censorship on Weibo\\Have been censored on Weibo\\Have used anti-censorship strategies \end{tabular} \\ \hline
\end{tabular}%
}
	\vspace{-0.3cm}
\end{table}

\begin{table}[htbp]
\centering
\caption{The correlation of three metrics of censorship and self-censorship (**means correlation is significant at the $p=0.01$ level).}\label{tab-index}
\begin{tabular}{cccc}
\hline
    & $I$     & $N$     & $S$     \\ \hline
$I$ & 1       & 0.511** & 0.396** \\
$N$ & 0.511** & 1       & 0.452** \\
$S$ & 0.396** & 0.452** & 1       \\ \hline
\end{tabular}%
	\vspace{-0.3cm}
\end{table}

\subsubsection{Mediation Model}
To test the relationships among perceived necessity of self-censorship, impact of self-censorship, and support for censorship, we follow MacKinnon's four-step procedure to establish the mediation effect~\cite{mackinnon2012introduction}, which requires (a) a significant association between necessity and support, (b) a significant association between necessity and impact, (c) a significant association between impact and support while controlling for necessity, and (d) a significant coefficient for the indirect path between necessity and support through impact. The bias-corrected percentile bootstrap method determines whether the last condition is satisfied.

\begin{figure}[htb]
	\centering
	\includegraphics[width=0.99\linewidth]{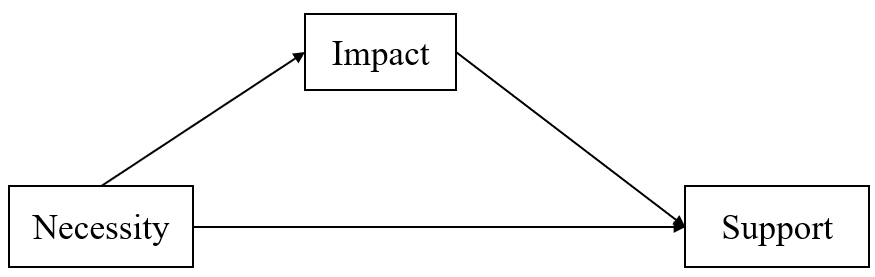}
	\vspace{-0.5cm}
	\caption{Mediation model constructed by three metrics of censorship and self-censorship: the perceived necessity of self-censorship, the impact of self-censorship on users' expression desire and mood, and the support for censorship.}
	\label{fig:4}
	\vspace{-0.4cm}
\end{figure}

Multiple regression analysis shows that, in the first step, necessity is significantly associated with support, $b=0.452, p<0.001$ (see Model 1 of Table~\ref{tab:my-table}). In the second step, necessity is significantly associated with impact,  $b=0.396, p<0.001$ (see Model 2 of Table~\ref{tab:my-table}). In the third step, when this study controls for necessity, impact is significantly associated with support, $b=0.296, p<0.001$ (see Model 3 of Table~\ref{tab:my-table}). Finally, the bias-corrected percentile bootstrap method indicats that the indirect effect of necessity on support through impact is significant, $ab=0.156$, $SE=0.023$, $95\% CI =[0.113, 0.202]$. The mediation effect accounted for 22.9\% of the total effect. Overall, necessity will increase impact, which in turn will increase support. In other words, impact will mediate the link between necessity and support (Figure~\ref{fig:4} shows the mediation effect).

\begin{table}[htb]
\centering
\caption{The mediation effect of necessity on support.}
\begin{tabular}{cllllllllllll}
\hline
 &
  \multicolumn{4}{l}{Model 1 ($S$)} &
  \multicolumn{4}{l}{Model 2 ($I$)} &
  \multicolumn{4}{l}{Model 3 ($S$)} \\ \cline{2-13} 
Predictors &
  \multicolumn{2}{l}{$b$} &
  \multicolumn{2}{l}{$t$} &
  \multicolumn{2}{l}{$b$} &
  \multicolumn{2}{l}{$t$} &
  \multicolumn{2}{l}{$b$} &
  \multicolumn{2}{l}{$t$} \\ \hline
$N$ &
  \multicolumn{2}{l}{0.452} &
  \multicolumn{2}{l}{11.564**} &
  \multicolumn{2}{l}{0.396} &
  \multicolumn{2}{l}{9.849**} &
  \multicolumn{2}{l}{0.296} &
  \multicolumn{2}{l}{7.596**} \\
$I$ &
  \multicolumn{2}{l}{} &
  \multicolumn{2}{l}{} &
  \multicolumn{2}{l}{} &
  \multicolumn{2}{l}{} &
  \multicolumn{2}{l}{0.394} &
  \multicolumn{2}{l}{10.118**} \\
$R^2$ &
  \multicolumn{2}{l}{0.204} &
  \multicolumn{2}{l}{} &
  \multicolumn{2}{l}{0.157} &
  \multicolumn{2}{l}{} &
  \multicolumn{2}{l}{0.335} &
  \multicolumn{2}{l}{} \\
$F$ &
  \multicolumn{2}{l}{133.726**} &
  \multicolumn{2}{l}{} &
  \multicolumn{2}{l}{97.001**} &
  \multicolumn{2}{l}{} &
  \multicolumn{2}{l}{131.063**} &
  \multicolumn{2}{l}{} \\ \hline
\end{tabular}
\label{tab:my-table}
	\vspace{-0.1cm}
\end{table}

\section{Discussion}

Social media censorship, including those facilitated by AI, is found prevalent in social media, especially in authoritarian regimes \cite{censorship:xu}. It greatly negates the freedom of speech. In this paper, we reveal the difficulties of expressing in social media in China from a user's perspective, updating existing findings (e.g., \cite{Chinese:Twitter}). We further explore the relationship between three intertwined concepts, i.e., censorship, self-censorship, and anti-censorship. Below, we reflect on main results and practical implications for more democratic social media in authoritarian regimes, and address limitations and future work in the end.   

\subsection{Recap of Findings}
Self-censorship is common on Chinese social media. Out of 523 valid samples, 90.4\% of the respondents acknowledged more or less practicing self-censorship. There was a consensus regarding the effect of self-censorship on people's willingness to express themselves and mood, which was not reported in previous literature \cite{self:censorship,media:censorship}. Common motivations for self-censorship included bypassing the censorship, avoiding being deleted, avoiding account suspension, avoiding being summoned by the police, etc. More politically sensitive topics, which are reported to be more susceptible to censorship \cite{deletion}, were more often self-censored by our respondents, echoing with prior research \cite{media:censorship}. Some examples included discussions of current events and the covid. Users also paid more attention to self-censorship when reposting than commenting.

79\% of our respondents had been censored at least once. While there is not an estimation of the ratio of censored Weibo posts, like that in the Turkish context \cite{turkey}, we provided a number in our limited sample. The consequences of censorship ranged from being deleted to being summoned by the police and even being put into prison \cite{prison}. Three anti-censorship strategies perceived as most effective to bypass the censorship infrastructure were satire, acronym, and homophones.

To examine the relationship between censorship and self-censorship, we defined three metrics, namely, perceived necessity of self-censorship, impact of self-censorship on users' expression desire and mood, and support for censorship. We then conducted correlation analysis, hypothesis testing, and regression analysis to reveal the influence factors that had a significant correlation with these three metrics. Finally, we conducted a regression analysis on these three metrics and established a mediation model to measure their relationship. Perceived necessity of self-censorship turns out to increase the impact of self-censorship on users' expression desire and mood, which in turn will increase support for censorship. A possible explanation is that those who deem self-censorship as necessary are more likely to practice self-censorship, and are thus more likely to be affected by it in terms of expression desire and mood. Those who deem self-censorship as necessary also tend to solve the issues brought by censorship from the end of themselves, instead of questioning the censorship infrastructure. Thus they are more likely to support censorship.


\subsection{Implications for Research and Design}
Several design and research implications can be drawn from our survey results.

Firstly, censorship \cite{censorship:2}, self-censorship \cite{media:censorship} and anti-censorship \cite{censorship:5} are not independent of each other. Due to the existence of the censorship infrastructure, Weibo users have adopted a number of anti-censorship strategies to defend their expression rights. It is also because of the censorship infrastructure that Weibo users have to practice self-censorship in order to freely express themselves and to avoid unnecessary trouble. These intertwined factors should be addressed together instead of separately in censorship research. Hereby we present an example by building a mediation model of censorship and self-censorship.

Secondly, it is increasingly hard to maintain a democratic and free social media environment in China. People are censored for various topics, especially politically sensitive ones such as current events and pandemic, and in different scenarios, i.e., posting, reposting, and commenting. In a blunting manner~\cite{blunting}, users may choose to not talk about these topics to avoid trouble. In a monitoring manner, users may proactively think of ways to resist the censorship infrastructure, e.g., English, metaphors, sarcasm, etc., as shown in our responses. On July 13, 2022, Weibo issued an announcement stating that it would carry out centralized rectification of illegal behaviors that used homophonic words, variant words and other typos to publish and disseminate ``bad information''~\cite{Typos}. As a result, the anti-censorship strategies used by our respondents will soon or later be deemed illegal. To bypass the censorship infrastructure which is intrinsically imposed by the government, a decentralized platform which is free of centralized regulation might be a rescue \cite{nft}.


\subsection{Limitations and Future Work}
The main limitation of our survey study is the skewed sampling, as 80\% of the respondents are female. The results may not generalize to a larger population with a different demographic. Nevertheless, our results reveal novel insights into this specific subset of Weibo users. Future work could consider conducting a larger-scale survey study to obtain more generalizable results, or conducting qualitative interviews to gain more in-depth insights of people's attitudes toward censorship.

\section{Conclusion}
In this paper, we conducted a survey study to understand the current landscape of Chinese social media in terms of censorship, self-censorship, and anti-censorship. People's attitudes toward censorship, self-censorship practices, and anti-censorship strategies are revealed through descriptive statistics, updating findings in previous research. A mediation model is further established to understand the relationship between censorship and self-censorship. Based on the results, we suggest a shift in censorship research, focusing on the relationship between censorship, self-censorship, and anti-censorship, and potential ways of democratizing social media in an authoritarian regime. 

\section*{Acknowledgements}
 We would like to thank the respondents who contributed to the survey data, and those who helped us distribute the survey. We also thank the anonymous reviewers for their good words and insightful feedback. We further thank Yifei Wang, Chuanli Xia, Zijie Shao, among many others, who provided useful feedback in the early stage of the study.

%
%
%
\bibliographystyle{splncs04}
\bibliography{mybibliography}

\end{document}